# High density THz frequency comb produced by coherent synchrotron radiation

S. Tammaro[a,c], O. Pirali[a,b], P. Roy[a], J.-F. Lampin[d], G. Ducournau[d], A. Cuisset[c], F. Hindle[c] , and G. Mouret[c]

[a] AILES Beamline, Synchrotron SOLEIL, l'Orme des Merisiers, Saint-Aubin, 91192 Gif-sur-Yvette cedex, France

[b] Institut des Sciences Moléculaires d'Orsay, UMR8214 CNRS – Université Paris-Sud, Bât. 210, 91405 Orsay cedex, France

[c] Laboratoire de Physico-Chimie de l'Atmosphère, Université du Littoral Côte d'Opale, 189A Avenue Maurice Schumann, 59140 Dunkerque, France

[d] Institut d'Electronique de Microélectronique et de Nanotechnologie, UMR8520 CNRS – Université de Lille 1, Avenue Poincaré-Cité Scientifique CS 60069, 59652 Villeneuve d'Ascq, France

**Frequency combs (FC) have radically changed the landscape of frequency metrology and high-resolution spectroscopy investigations extending tremendously the achievable resolution while increasing signal to noise ratio. Initially developed in the visible and near-IR spectral regions[1], the use of FC has been expanded to mid-IR[2], extreme ultra-violet[3] and X-ray[4]. Significant effort is presently dedicated to the generation of FC at THz frequencies. One solution based on converting a stabilized optical frequency comb using a photoconductive terahertz emitter, remains hampered by the low available THz power[5]. Another approach is based on active mode locked THz quantum-cascade-lasers providing intense FC over a relatively limited spectral extension[6,7,8]. Alternatively, here we show that dense powerful THz FC is generated over one decade of frequency by coherent synchrotron radiation (CSR). In this mode, the entire ring behaves in a similar fashion to a THz resonator wherein electron bunches emit powerful THz pulses quasi-synchronously. This FC presents unprecedented spectral characteristics and opens new perspectives for metrology investigations.**

As an alternative to laser based experiments and electronic generation, broadband THz radiation emitted from short electron bunches is now used for high-resolution spectroscopy and spectro-microscopy applications. In the so-called low-$\alpha$ mode, intense THz power is obtained by reducing the electron bunch length to produce a phase coherent synchrotron radiation[9]. Since the first demonstration, such radiation has been developed in several synchrotron facilities over the world. It has provided unprecedented power in the region 0.1 THz -1 THz, five orders of magnitude brighter than standard incoherent synchrotron radiation[10]. Super-Radiance effects obtained from bunch to bunch interference[11] further enhance specific frequencies (1/$T_{bunch}$ where $T_{bunch}$ is the time interval between successive bunches) of the THz CSR. Here, using a dedicated high-performance sub-THz heterodyne receiver (see Methods), we reveal for the first time that THz CSR is an intense, stable (in both frequency and amplitude) zero offset FC covering a wide spectral range 0.1-1THz. This revolutionizes the received opinion that considers and uses CSR as a continuum[10]. Moreover we confirm its potential use for spectroscopy with a high resolution absorption measurement.

In this work, the CSR spectral structure has been probed by an existing Bruker Fourier Transform Infrared (FTIR) instrument with an ultimate resolution of 30MHz and the proposed heterodyne receiver shown in figure 1. The receiver, capable of operation in a large spectral region with a bandwidth of about 10 GHz, it provides unprecedented spectral resolution, high signal to noise ratio and excellent frequency accuracy. Using both spectrometers we are able to detect the fine spectral features of the CSR. In particular, we identified two main ultra-stable FC sequences in both regions studied (200±10 GHz and 400±10 GHz). The FC sequences are illustrated on Figure 2 with a progressively finer frequency scale. The upper panel of this figure shows the CSR spectrum as recorded by the FTIR spectrometer at the highest possible spectral resolution (maximum optical path difference of 8.82 m, corresponding to a resolution of 30 MHz). The full CSR range convoluted by FTIR response is shown in figure 2a. A detailed view around 600 GHz reveals a 352 MHz FC superimposed on the broad THz emission, figure 2b, as previously observed elsewhere[11]. The lower panel shows the heterodyne analysis of the CSR. Figure 2c shows 3 GHz of heterodyne spectrum at 200 GHz, the 352 MHz FC originating from the bunch to bunch repetition is also clearly resolved. This FC is the signature of a degree of coherence amongst the electron bunches[12] and is designated as super radiance emission. Expanding the frequency scale reveals a second FC composed of sharp teeth regularly spaced by 846 kHz, figure 2d. This comb is indeed related to the very stable revolution period of the electron bunches in the storage ring (1.18 µs). It produces a spectrally dense THz FC with more than $10^6$ components covering the THz range from 0.1 to 1 THz. The observed structure as shown in figure 2d is composed of the lower and upper frequency band contributions as expected for heterodyne detection technique. Aliasing of the FC results in the observation of a doubled FC structure whose frequency separation is dependent on the local oscillator (LO) frequency. In our case the LO frequency was selected to distinguish the two components and resulted in a separation of 100 kHz, figure 2e. The heterodyne analysis allowed the power of an individual CSR line around 400 GHz to be estimated at -77dBm (i.e. 20pW) (see Methods). The CSR spectral power density is 3 orders of magnitude greater than can be achieved from a classical dipolar antenna used in a context of THz Time Domain[5]. The detailed examination of any individual tooth, inset 2e, indicates an even narrower structure: two principal components separated by 1.3 kHz, each with a FWHM systematically lower than 500Hz. The substructure is interpreted as a consequence of the kHz range amplitude modulation attributed to low frequency instabilities of the electron bunches[13]. In general, FC modes are described as multiples of a reference frequency ($f_R$) shifted by an offset ($f_0$): $f_{FC} = n \times f_R + f_0$ where $n$ is an integer. The offset frequency was assessed by measuring the revolution frequency with a counter (CNT 90, Pendulum). Over a period of 20 seconds, the revolution frequency is determined with a typical uncertainty of 30µHz. The comb mode center frequencies were determined by the spectral analysis of the IF signal. In excess of 50 modes were examined, each was systematically found to be an exact multiple of the revolution frequency with a confidence interval of 1 Hz. Unlike optical FC, the CSR FC is therefore considered to be offset free.

CSR FC with heterodyne receiver is a powerful combination capable of excellent resolving power, frequency metrology and high dynamic range of the amplitude. To illustrate such properties, high resolution spectroscopy measurements were undertaken. We recorded the absorption spectrum of acetonitrile ($CH_3CN$) as the K series of this symmetric top molecule are particularly well suited to test the frequency accuracy and the dynamic range of the CSR FC. A 1.3m long absorption gas cell was inserted into the CSR path. Figure 3 shows a small section of the acetonitrile spectrum obtained in 65 seconds using the CSR FC with the LO at 202 GHz. The FC gives rise to a discrete spectrum built from the harmonics of the repetition rate of 846 kHz (violet vertical lines). The complete K structure of the R(10) transition of $CH_3CN$ is clearly resolved (violet profile, as explained in the associated content). This is in strong contrast with the simulated spectrum corresponding to the 30 MHz resolution of the IFS125 interferometer (red profile), only displaying a broad unresolved feature. In particular, the THz FC allows the doublet K=0 and K=1 separated by 3.9 MHz ($1.3 \times 10^{-4}$ $cm^{-1}$) to be resolved. The fit of the 5 individual absorption lines with Voigt profiles provides their central frequencies, which are in excellent agreement with reference data (RMS < 80 kHz) (see Methods). The accuracy obtained is comparable with frequency multiplication techniques currently employed in this frequency range. The dynamic range of the amplitude of the CSR lines is systematically in excess of 35 dB with the strongest lines exhibiting over 70 dB.

The FC frequency stability allows for room temperature heterodyne fast detection revealing new properties of synchrotron radiation. These promising results clearly encourage the development of receivers for ultra-high resolution over broader frequency ranges. This is an enabling approach for new high resolution THz time-resolved experiments. Amongst highly desirable new applications are ultrafast monitoring of reaction kinetics, ultra-high resolution atomic and molecular spectroscopy[14], and spatio-temporal dynamics of relativistic electron bunches.

# METHODS

**Coherent Synchrotron Radiation**

The complete sequences of intense THz FC combs were observed in the low $\alpha$ coherent mode characterized by a 4.8ps RMS bunch length and corresponding to a compaction factor $\alpha = \alpha_0/25$, with $\alpha_0 = 1.7\times10^{-5}$ the nominal value[13]. The ring was injected with 208 electron bunches distributed on one half of the SOLEIL ring and by one isolated bunch in the centre of the other half. In this operation mode the total current was about 16mA with periodic refilling in top-up mode. Two principal periods characterize the temporal radiation distribution: i) the 208 bunches produce a 2.8ns periodic pulsed radiation train and ii) the revolution period in the ring is about 1.18$\mu$s.

**Heterodyne detection**

To investigate the spectral structure of the THz CSR, we constructed a high sensitivity coherent electronic receiver[15]. Heterodyne detection is very convenient for high resolution spectral analysis with a high dynamic range. The operation is based on a frequency translation of the analyzed spectra by electrical mixing between a known reference frequency (local oscillator) and the signal to be analyzed. The mixing element is a Schottky-based sub-harmonic-mixer (SHM), fed by the local oscillator at mW level (continuous wave mode), this power being required to operate the Schottky device in non-linear regime. Two different SHM mixers WR5.1 (140-220 GHz) and WR 2.2 (325-500 GHz) are used for operation at 200 GHz and 400 GHz respectively. The LO is generated by an electronic frequency multiplication chain driven by a Rohde & Schwarz SMF100A synthesizer. This provides equivalent frequency multiplication factors of

×12 and ×24 respectively. The IF signal is amplified with two base-band amplifiers, providing a total gain of 45 dB over a bandwidth of 500 MHz. The down-converted CSR signal is analysed by a 3.6 GHz Agilent MXA Signal Analyzer N9020A.

**Power measurements**

From IF signal spectral analysis, the individual CSR lines around 400 GHz display a typical signal strength of -40 dBm. Taking into account the 45 dB IF gain and a 8 dB conversion loss of SHM, the corresponding power of CSR lines should be around -77 dBm (20 pW). The integrated THz power of the CSR was measured to be 60.5 $\mu$W by a Gentec THz 5B-BL-DZ pyroelectric detector working at room temperature. The beam was filtered with a black polyethylene film in order to avoid any effect of visible or mid-IR radiation. The ratio of the integrated power to the power of one line is then equal to about $3\times10^6$. This gives a rough estimation of the number of lines assuming that all lines of the FC have the same intensity. This value is compatible with the fact that in a 1 THz range there should be about $1.2\times10^6$ lines separated by 846 kHz. The spectral power density is hence calculated to be 23.6 nW/GHz.

**Spectroscopic analysis on acetonitrile**

The IF spectra are processed to separate the lower and upper frequency band contributions. Once all the peaks in the spectra have been identified the FC spacing of 846 kHz is used to isolate each FC component. In figure 3, the normalized absorbance of $CH_3CN$ measured with the THz FC was simulated with Voigt profiles. The Doppler line width (Gaussian contributions) of the 5 components has been fixed to a calculated value at room temperature of 0.4 MHz (FWHM). The center frequencies and the collisional line widths (Lorentzian contributions) were fitted for K = 0

- 5 components of the R(10) transitions. In table 1, the fitted frequencies $f_{fitted}$ are compared with the frequencies $f_{calc}$ listed in the JPL database[16] and calculated from the latest combined fit reported by H. S. P. Müller *et al.*[17]. The r.m.s determined from the ($f_{fitted}$ - $f_{calc}$) values is evaluated to be to 79 kHz. The collisional line widths were fitted to 3.4 MHz (FWHM) for the 5 components (no K dependence of the R(10) self-broadening was observed). At a pressure of 50 μbar a collisional linewidth of around 4.5 MHz (FWHM) may be estimated for the R(10) transitions from $CH_3CN$ self broadening measurements[18, 19]

| K | $f_{fitted}$ (MHz) | $f_{calc}$ (MHz) | $f_{fitted}$ - $f_{calc}$ (MHz) | 100×($f_{fitted}$ - $f_{calc}$)/ $f_{calc}$ |
|---|---|---|---|---|
| 0 | 202355.39 | 202355.51 | -0.12 | 5.93E-05 |
| 1 | 202351.73 | 202351.61 | 0.12 | 5.93E-05 |
| 2 | 202339.88 | 202339.92 | -0.04 | 1.97E-05 |
| 3 | 202320.42 | 202320.44 | -0.02 | 9.89E-06 |
| 4 | 202293.19 | 202293.18 | 0.01 | 4.94E-06 |

Table 1: fitted frequencies compared with the frequencies listed in the JPL database[16]

# REFERENCES

1. Udem, Th., Holzwarth, H., Hänsch, T. W., Optical frequency metrology. *Nature* **416**, 233-237 (2002).

2. Schliesser, A., Picqué, N., Hänsch, T. W., Mid-infrared frequency combs. *Nature Photon.* **6**, 440 (2012).

3. Zinkstok, R. Th., Witte, S., Ubachs, W., Hogervorst, W., Eikema, K. S. E., Frequency comb laser spectroscopy in the vacuum-ultraviolet region. *Physical Review A* **73**, 061801 (2006).

4. Cavaletto, S. M. *et al.* Broadband high-resolution X-ray frequency combs. *Nature Photon.* **8**, 520-523 (2014).

5. Tani, M., Matsuura, S., Sakai, K., Nakashima, S. I., Emission characteristics of photoconductive antennas based on low-temperature-grown GaAs and semi-insulating GaAs. *Applied Optics* **36,** 7853-7859 (1997).

6. Barbieri, S. *et al.* Coherent sampling of active mode-locked terahertz quantum cascade lasers and frequency synthesis. *Nature Photon.* **5**, 306-313 (2011).

7. Maysonnave, J. *et al.* Mode-locking of a terahertz laser by direct phase synchronization. *Optics Express* **20**, 20855-62 (2012).

8. Burghoff, D. *et al.* Terahertz laser frequency combs. *Nature Photon.* **8,** 462-467 (2014) .

9. Abo-Bakr, M. *et al.* Brilliant, Coherent Far-Infrared (THz) Synchrotron Radiation. *Physical Review Letters* **90**, 094801 (2003).

10. Barros, J. *et al.* Coherent synchrotron radiation for broadband terahertz spectroscopy. *Review of Scientific Instruments* **84**, 033102 (2013).

11. Billinghurst, B.E. *et al.* Observation of superradiant synchrotron radiation in the terahertz region. *Physical Review Special Topics* **16,** 060702 (2013).

12. Shibata, Y. *et al.* Observation of interference between coherent synchrotron radiation from periodic bunches. *Physical Review A* **44,** R3445-R3448 (1991).

13. Evain, C. *et al*. Spatio-temporal dynamics of relativistic electron bunches during the micro-bunching instability in storage rings. *EPL* **98**, 40006 (2012).

14. Bartalini, S. *et al.* Frequency-Comb-Assisted Terahertz Quantum Cascade Laser Spectroscopy. *Phys.Rev. X* **4,** 021006 (2014).
15. Maas, S.A., Microwave Mixers, Artech House (1992).
16. Pickett, H. M. *et al.,* Submillimeter, Millimeter, and Microwave Spectral Line Catalog. *J. Quant. Spectrosc. & Rad. Transfer* **60,** 883-890 (1998).
17. Müller, H. S. P., Drouin, B. J., Pearson, J. C., Rotational spectra of isotopic species of methyl cyanide, $CH_3CN$, in their ground vibrational states up to terahertz frequencies. *Astron. Astrophys.* **506,** 1487 (2009).
18. Rinsland, C. P. *et al*. Multispectrum analysis of the $\nu_4$ band of $CH_3CN$: Positions, intensities, self- and $N_2$-broadening, and pressure-induced shifts. *Journal of Quantitative Spectroscopy & Radiative Transfer*, **109,** 974-994 (2008).
19. Buffa, G. *et al*. Far infrared self broadening in methylcyanide. *Physical Review A*, **45**, 6443-6450 (1992).

**ACKNOWLEDGMENT**

This work was partly supported by the Université du Littoral Côte d'Opale in the framework of the « Bonus Qualité Recherche » and the synchrotron SOLEIL under the proposal 20131346. The authors would like to acknowledge the support of Jean-Blaise Brubach (Soleil), Laurent Manceron (Soleil) and Marc Fourmentin (LPCA).

**Author Contributions**

OP and GM wrote the synchrotron beamtime proposal and supervised the project. Experiments were performed by ST, OP, JFL, GD, FH and GM. The analysis was made by ST, PR, JFL, GD, AC, FH and GM. All the authors contributed equally to the preparation of the manuscript.

**Notes**

The authors declare no competing financial interest.

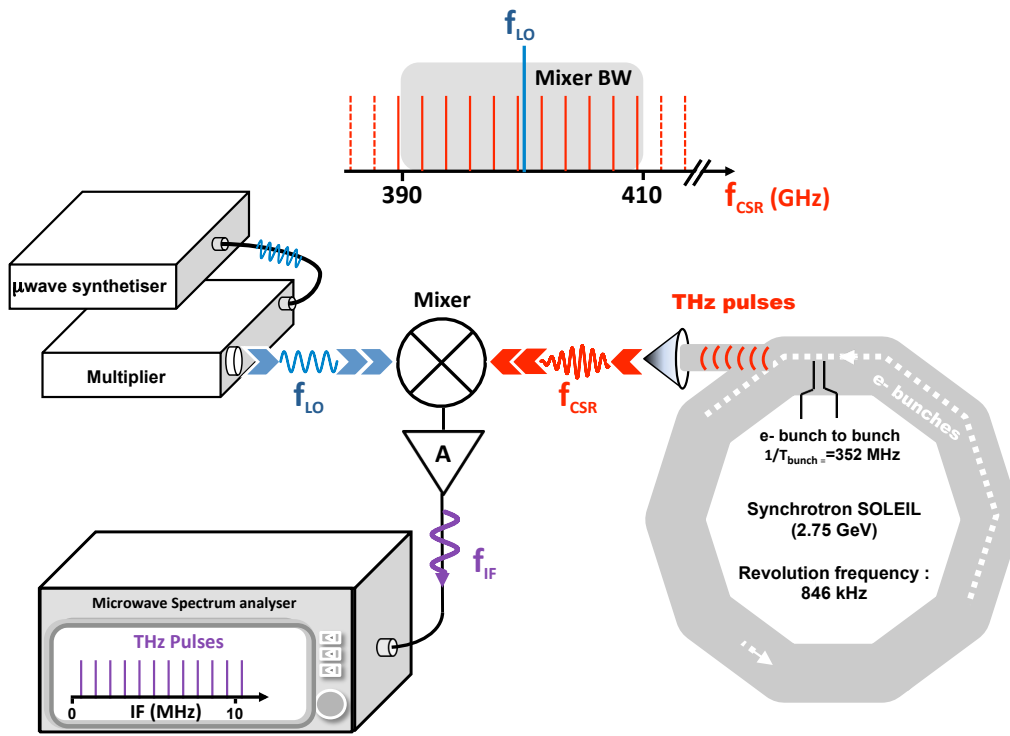

**Figure 1: CSR heterodyne detection schematic.** Short bunches of relativistic electrons circulating in the SOLEIL storage ring emit CSR in the 0.1-1 THz range. The bunch to bunch frequency $1/T_{bunch}$ is 352 MHz, while the storage ring revolution frequency is 846 kHz (see Methods). The THz CSR pulses are mixed with the monochromatic radiation ($f_{LO}$) from a local oscillator (microwave synthesizer and multiplier) creating an electronic signal at the intermediate frequency (IF): $f_{IF} = |f_{LO}-f_{CSR}|$ within the mixer bandwidth. The resulting signal is amplified by the IF amplifier A and analyzed using a microwave spectrum analyzer. The mixing produces a frequency down conversion of $f_{CSR}$ into the microwave range, and an aliasing of the spectrum due to the superposition of the upper ($f_{CSR} > f_{LO}$) and the lower ($f_{CSR} < f_{LO}$) frequency bands.

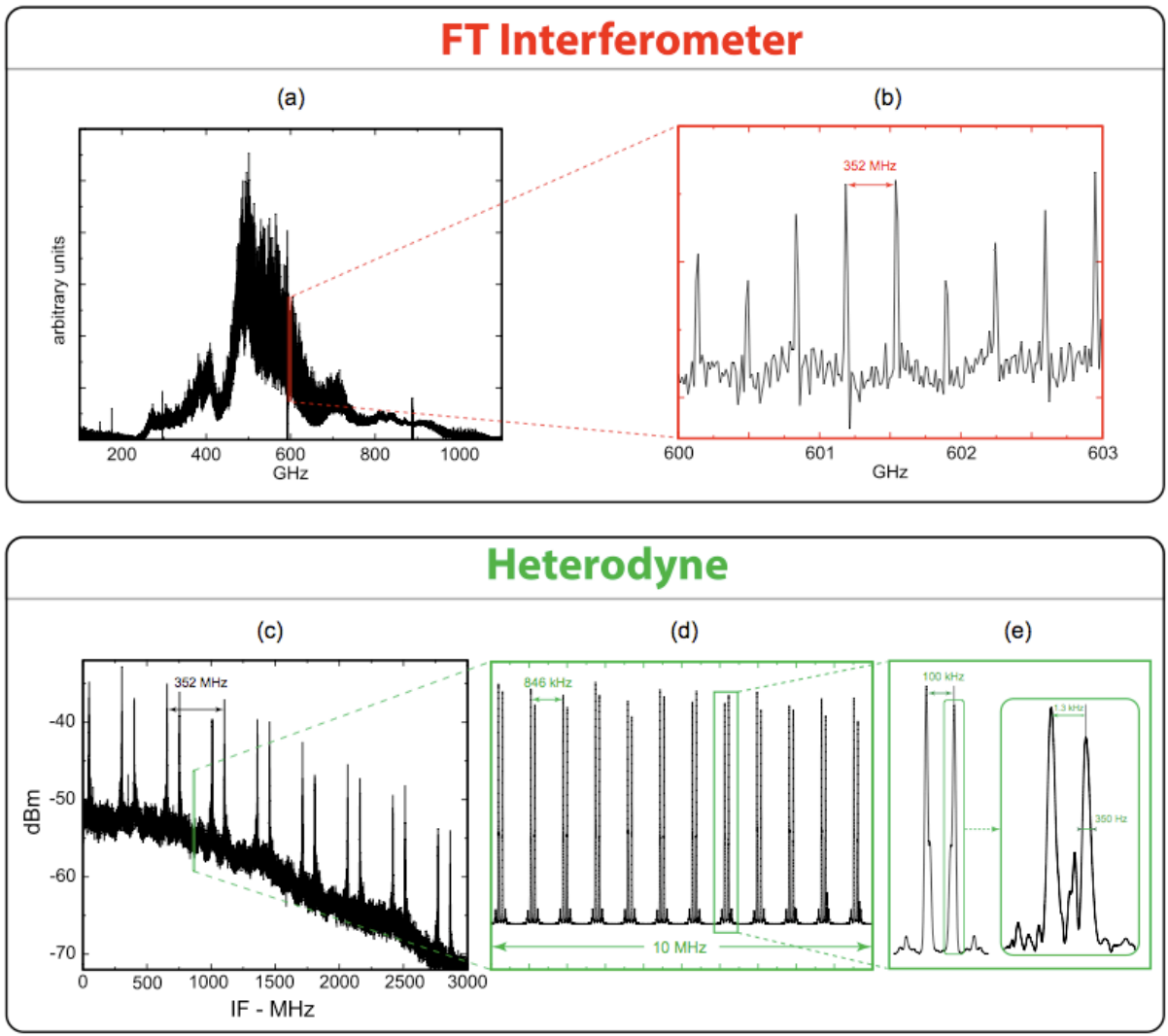

**Figure 2: Typical CSR FC structure recorded using the FTIR spectrometer (upper pane) and using the heterodyne receiver (lower pane).** All amplitude scales are linear unless stated otherwise. Heterodyne frequency scales are established as a function of the intermediate frequency. 2a: full range of the CSR measured by the FTIR spectrometer with a resolution of 30 MHz; the low-frequency response is limited by the optical elements of the instrument. 2b: zoom of 3 GHz revealing the 352 MHz FC corresponding to $1/T_{bunch}$. 2c: heterodyne spectrum obtained by mixing CSR with $f_{LO}$=200 GHz. The FC at 352 MHz is also easily distinguished. 2d: the second FC composed of sharp teeth regularly spaced by 846 kHz is revealed by an expanded frequency scale. This second comb is produced by the very stable revolution period of

the electron bunches in the storage ring (1.18 µs). 2e: a zoom of a given dual comb structure caused by the aliasing of the upper and lower frequency bands. Inset 2e: the excellent resolving power of the frequency scale allows the detailed examination of a single tooth highlighting the ultrafine splitting due to low frequency instabilities of the electron bunches.

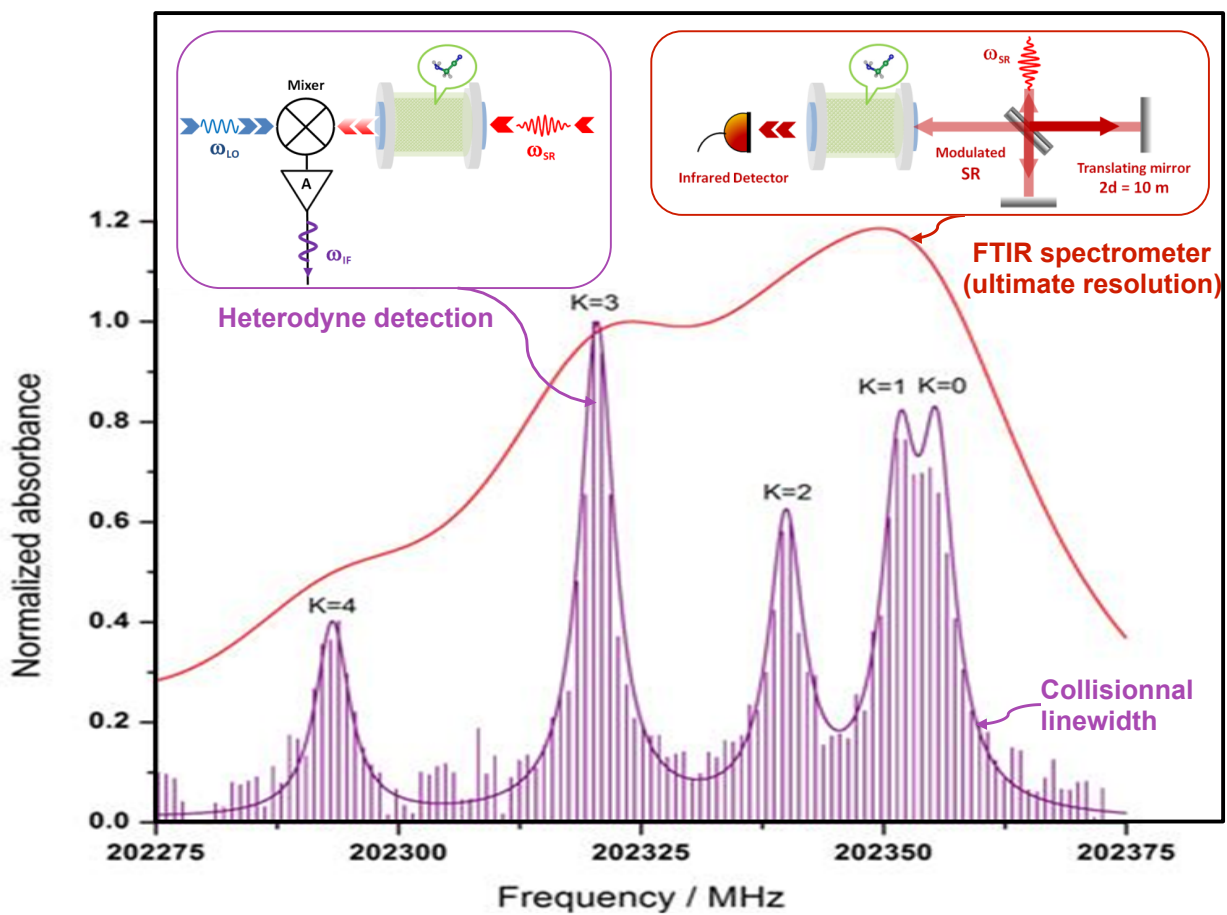

**Figure 3: Normalized absorbance of acetonitrile.** Normalized absorbance versus frequency showing the K structure of the pure rotational R(10) transition of acetonitrile recorded with the heterodyne receiver, $f_{LO}$=202.3 GHz (violet vertical lines). The gas pressure was set to 50 µbar. The violet solid line is the result of the fit of the 5 individual lines using Voigt profiles (see Methods) while the red curve corresponds to a simulation at the maximum possible resolution (30 MHz) of commercially available FT spectrometers. The instrument configurations are schematically represented in the insets. Left: heterodyne measurement, the synchrotron radiation transmitted through a cell filled with gas is incident on a mixer simultaneously with the radiation from a local oscillator. Right: classical absorbance measurement based on FTIR method, the interferometer modulated radiation passes through a gas containing cell before arriving at the detector.